\documentclass[journal,10pt]{IEEEtran}

\def\formula{\fontsize{10}{12}\normalfont\rmfamily\selectfont}

\usepackage{latexsym}
\usepackage{epsfig}
\usepackage{amsmath}
\usepackage{amsfonts}
\usepackage{amssymb}
\usepackage{wrapfig}
\usepackage{float}
\usepackage{bm}

\newtheorem{theorem}{Theorem}

\newtheorem{lemma}{Lemma}
\newtheorem{proposition}{Proposition}
\newtheorem{remark}{Remark}

\begin{document}

\title{Q-CSMA: Queue-Length Based CSMA/CA Algorithms for Achieving Maximum Throughput and Low Delay in Wireless Networks\thanks{Research supported by NSF Grants 07-21286, 05-19691, 03-25673, ARO MURI Subcontracts, AFOSR Grant
FA-9550-08-1-0432 and DTRA Grant HDTRA1-08-1-0016.}}

\author{\authorblockN{Jian Ni, Bo (Rambo) Tan, and R. Srikant\\}
\authorblockA{Coordinated Science Laboratory and Department of Electrical and Computer Engineering\\
University of Illinois at Urbana-Champaign, Urbana, IL 61801, USA
\\ Email: \{jianni, botan2, rsrikant\}@illinois.edu} }

\maketitle \thispagestyle{empty}

\begin{abstract}

Recently, it has been shown that CSMA-type random access algorithms can achieve the maximum possible
throughput in ad hoc wireless networks. However, these algorithms assume an idealized continuous-time CSMA
protocol where collisions can never occur. In addition, simulation results indicate that the delay
performance of these algorithms can be quite bad. On the other hand, although some simple heuristics (such as
distributed approximations of greedy maximal scheduling) can yield much better delay performance for a large
set of arrival rates, they may only achieve a fraction of the capacity region in general. In this paper, we
propose a discrete-time version of the CSMA algorithm. Central to our results is a discrete-time distributed
randomized algorithm which is based on a generalization of the so-called Glauber dynamics from statistical
physics, where multiple links are allowed to update their states in a single time slot. The algorithm
generates collision-free transmission schedules while explicitly taking collisions into account during  the
control phase of the protocol, thus relaxing the perfect CSMA assumption. More importantly, the algorithm
allows us to incorporate mechanisms which lead to very good delay performance while retaining the
throughput-optimality property. It also resolves the hidden and exposed terminal problems associated with
wireless networks.

\end{abstract}

\section{Introduction}   \label{sec:introduction}

For wireless networks with limited resources, efficient resource allocation and optimization (e.g., power
control, link scheduling, routing, congestion control) play an important role in achieving high performance
and providing satisfactory quality-of-service (QoS). In this paper, we study \emph{link scheduling} (or
\emph{Media Access Control}, MAC) for wireless networks, where the links (node pairs) may not be able to
transmit simultaneously due to transceiver constraints and radio interference. A \emph{scheduling algorithm}
(or \emph{MAC protocol}) decides which links can transmit data at each time instant so that no two active
links interfere with each other.

The performance metrics of interest in this paper are \emph{throughput} and \emph{delay}. The throughput
performance of a scheduling algorithm is often characterized by the largest set of arrival rates under which
the algorithm can keep the queues in the network stable. The delay performance of a scheduling algorithm can
be characterized by the average delay experienced by the packets transmitted in the network. Since many
wireless network applications have stringent bandwidth and delay requirements, designing high-performance
scheduling algorithms to achieve maximum possible throughput and low delay is of great importance, which is
the main objective of this paper. We also want the scheduling algorithms to be distributed and have low
complexity/overhead, since in many wireless networks there is no centralized entity and the resources at the
nodes are very limited.

It is well known that the queue-length based \emph{Maximum Weight Scheduling} (MWS) algorithm is
\emph{throughput-optimal} \cite{taseph92}, in the sense that it can stabilize the network queues for all
arrival rates in the capacity region of the network (without explicitly knowing the arrival rates). However,
for general interference models MWS requires the network to solve a complex combinatorial optimization
problem in each time slot and hence, is not implementable in practice.

\emph{Maximal scheduling} is a low-complexity alternative to MWS but it may only achieve a small fraction of
the capacity region \cite{chakarsar05,wusriper07}. \emph{Greedy Maximal Scheduling} (GMS), also known as
\emph{Longest-Queue-First} (LQF) Scheduling, is another natural low-complexity alternative to MWS which has
been observed to achieve very good throughput and delay performance in a variety of wireless network
scenarios. GMS proceeds in a greedy manner by sequentially scheduling a link with the longest queue and
disabling all its interfering links. It was shown in \cite{dimwal06} that if the network satisfies the
so-called \emph{local-pooling} condition, then GMS is throughput-optimal; but for networks with general
topology GMS may only achieve a fraction of the capacity region \cite{joolinshr08,lecnisri09,zusbrzmod08}.
Moreover, while the computational complexity of GMS is low, the signaling and time overhead of
decentralization of GMS can increase with the size of the network \cite{lecnisri09}.

Another class of scheduling algorithms are CSMA (\emph{Carrier Sense Multiple Access}) type random access
algorithms. Under CSMA, a node (sender of a link) will sense whether the channel is busy before it transmits
a packet. When the node detects that the channel is busy, it will wait for a random backoff time. Since
CSMA-type algorithms can be easily implemented in a distributed manner, they are widely used in practice
(e.g., the IEEE 802.11 MAC protocol). In \cite{boo87} the authors derived an analytical model to calculate
the throughput of a CSMA-type algorithm in multi-hop wireless networks. They showed that the Markov chain
describing the evolution of schedules has a product-form stationary distribution under an idealized
continuous-time CSMA protocol (which assumes zero propagation/sensing delay and no hidden terminals) where
collisions can never occur. Then the authors proposed a heuristic algorithm to select the CSMA parameters so
that the link service rates are equal to the link arrival rates which were assumed to be known. No proof was
given for the convergence of this algorithm. This model was used in \cite{wankar05} to study throughput and
fairness issues in wireless ad hoc networks. The insensitivity properties of such a CSMA algorithm have been
recently studied in~\cite{liekaileuwon07}.

Based on the results in \cite{boo87,wankar05,liekaileuwon07}, a distributed algorithm was developed in
\cite{jiawal08} to adaptively choose the CSMA parameters to meet the traffic demand without explicitly
knowing the arrival rates. The results in \cite{jiawal08} make a time-scale separation assumption, whereby
the CSMA Markov chain converges to its steady-state distribution instantaneously compared to the time-scale
of adaptation of the CSMA parameters. Then the authors suggested that this time-scale separation assumption
can be justified using a stochastic-approximation type argument which was verified in
\cite{liuyipro08,jiawal08b}. Preliminary ideas for a related result was reported in \cite{rajsha08}, where
the authors study distributed algorithms for optical networks. But it is clear that their model also applies
to wireless networks with CSMA. In \cite{rajshashi08}, a slightly modified version of the algorithm proposed
in \cite{rajsha08} was shown to be throughput-optimal. The key idea in \cite{rajshashi08} is to choose the
link weights to be a specific function of the queue lengths to essentially separate the time scales of the
link weights and the CSMA dynamics. Further, the results in \cite{rajshashi08} assume that the
max-queue-length in the network is known which is estimated via a distributed message-passing procedure.
Similar modifications to 802.11 have been studied in \cite{akyetal08,warjanrhe09} where the back-pressure on
a link is used as the weight instead of the queue length.

While the results in our paper are most closely related to the works in \cite{boo87,jiawal08,rajshashi08}, we
also note important contributions in \cite{bormcdpro08,durthi06,mareryozd07,proyichi08} which make
connections between random access algorithms and stochastic loss networks.

Although the recent results on CSMA-type random access algorithms show throughput-optimality, simulation
results indicate that the delay performance of these algorithms can be quite bad and much worse than MWS and
GMS. Thus, one of our goals in this paper is to design \emph{distributed} scheduling algorithms that have
\emph{low complexity}, are \emph{provably throughput-optimal}, and have \emph{good delay performance}.
Towards this end, we design a discrete-time version of the CSMA random access algorithm.  It is based on a
generalization of the so-called Glauber dynamics from statistical physics, where multiple links are allowed
to update their states based on their queue lengths in a single time slot. Our algorithm generates
collision-free data transmission schedules while allowing for collisions during the control phase of the
protocol (as in the 802.11 MAC protocol), thus relaxing the perfect CSMA assumption of the algorithms studied
in \cite{boo87,jiawal08,rajshashi08}. Our approach to modeling collisions is different from the approaches in
\cite{jiawal09,liuyipro08}. In \cite{liuyipro08} the authors pointed out that, as the transmission
probabilities are made small and the transmission lengths are made large, their discrete-time model
approximates the continuous-time model with Poisson clocks, but it is difficult to quantify the throughput
difference between these two models. The algorithm in \cite{jiawal09} places upper bounds on the CSMA
parameters, while the loss in throughput due to this design choice is also hard to quantify. Instead, we
directly quantify the loss in throughput as the ratio of the duration of the control slot to the duration of
the data slot (see Remark~\ref{remark:overhead} in Section~\ref{sec:distributed}). More importantly, our
formulation allows us to incorporate delay-reduction mechanisms in the choice of schedules while retaining
the algorithm's throughput-optimality property. It also allows us to resolve the hidden and exposed terminal
problems associated with wireless networks \cite{MACAW}.

We organize the paper as follows. In Section~\ref{sec:model} we introduce the network model. In
Section~\ref{sec:basic} we present the basic scheduling algorithm and show that the (discrete-time) Markov
chain of the transmission schedules has a product-form distribution. In Section~\ref{sec:distributed} we
present a distributed implementation of the basic scheduling algorithm, called Q-CSMA (Queue-length based
CSMA/CA). In Section~\ref{sec:hybrid} we propose a hybrid Q-CSMA algorithm which combines Q-CSMA with a
distributed procedure that approximates GMS to achieve both maximum throughput and low delay. We evaluate the
performance of different scheduling algorithms via simulations in Section~\ref{sec:simulation}. The paper is
concluded in Section~\ref{sec:conclusion}.

\section{Network Model}   \label{sec:model}

We model a (single-channel) wireless network by a graph $G=(V, E)$, where $V$ is the set of nodes and $E$ is
the set of links. Nodes are wireless transmitters/receivers. There exists a directed link $(n,m)\in E$ if
node $m$ can hear the transmission of node $n$. We assume that if $(n,m)\in E$, then $(m,n)\in E$.

For any link $i\in E$, we use $\mathcal{C}(i)$ to denote the set of conflicting links (called \emph{conflict
set}) of $i$, i.e., $\mathcal{C}(i)$ is the set of links such that if any one of them is active, then link
$i$ cannot be active. The conflict set $\mathcal{C}(i)$ may include
\begin{itemize}
\item Links that share a common node with link $i$. This models the \emph{node-exclusive constraint} where
two links sharing a common node cannot be active simultaneously.

\item Links that will cause interference to link $i$ when transmitting. This models the \emph{radio
interference constraint} where two links that are close to each other cannot be active simultaneously.
\end{itemize}
We assume symmetry in the conflict set so that if $i\in \mathcal{C}(j)$ then $j\in \mathcal{C}(i)$.

We consider a time-slotted system. A \emph{feasible} (\emph{collision-free}) \emph{schedule} of $G=(V,E)$ is
a set of links that can be active at the same time according to the conflict set constraint, i.e., no two
links in a feasible schedule conflict with each other. We assume that all links have unit capacity, i.e., an
active link can transmit one packet in one time slot under a feasible schedule. Note that the results in this
paper can be readily extended to networks with arbitrary link capacities.

A schedule is represented by a vector $\mathbf x \in \{0,1\}^{|E|}$. The $i^{\rm th}$ element of $\mathbf x$
is equal to $1$ (i.e., $x_i=1$) if link $i$ is included in the schedule; $x_i=0$ otherwise. With a little bit
abuse of notation, we also treat $\mathbf x$ as a set and write $i\in \mathbf x$ if $x_i=1$. Note that a
feasible schedule $\mathbf x$ satisfies the following condition:
\begin{equation} \label{eq:feasibleschedule}
x_i + x_j \leq 1, \mbox{ for all } i \in E \mbox{ and } j \in \mathcal{C}(i).
\end{equation}
Let $\mathcal{M}$ be the set of all feasible schedules of the network.

A \emph{scheduling algorithm} is a procedure to decide which schedule to be used (i.e., which set of links to
be activated) in every time slot for data transmission. In this paper we focus on the MAC layer so we only
consider single-hop traffic. The \emph{capacity region} of the network is the set of all arrival rates
$\boldsymbol \lambda$ for which there exists a scheduling algorithm that can stabilize the queues, i.e., the
queues are bounded in some appropriate stochastic sense depending on the arrival model used. For the purposes
of this paper, we will assume that if the arrival process is stochastic, then the resulting queue length
process admits a Markovian description, in which case, stability refers to the positive recurrence of this
Markov chain. It is known (e.g., \cite{taseph92}) that the capacity region is given by
\begin{equation} \label{eq:capacity}
\Lambda = \{\boldsymbol{\lambda}\mbox{ }|\mbox{ }\boldsymbol{\lambda} \geq \mathbf{0} \mbox{ and } \exists
\boldsymbol \mu\in Co(\mathcal{M}),\boldsymbol{\lambda}<\boldsymbol{\mu}\},
\end{equation}
where $Co(\mathcal{M})$ is the \emph{convex hull} of the set of feasible schedules in $\mathcal M.$ When
dealing with vectors, inequalities are interpreted component-wise.

We say that a scheduling algorithm is \emph{throughput-optimal}, or achieves the \emph{maximum throughput},
if it can keep the network stable for all arrival rates in $\Lambda$.

\section{The Basic Scheduling Algorithm}   \label{sec:basic}

We divide each time slot $t$ into a \emph{control} slot and a \emph{data} slot. (Later, we will further
divide the control slot into control mini-slots.) The purpose of the control slot is to generate a
collision-free \emph{transmission schedule} $\mathbf x(t)\in \mathcal M$ used for data transmission in the
data slot. To achieve this, the network first selects a set of links that do not conflict with each other,
denoted by $\mathbf m(t)$. Note that these links also form a feasible schedule, but it is not the schedule
used for data transmission. We call $\mathbf m(t)$ the \emph{decision schedule} in time slot $t$.

Let $\mathcal{M}_0 \subseteq \mathcal M$ be the set of possible decision schedules. The network selects a
decision schedule according to a randomized procedure, i.e.,  it selects $\mathbf m(t) \in \mathcal{M}_0$
with positive probability $\alpha(\mathbf m(t))$, where $\sum_{\mathbf m(t)\in \mathcal M_0} \alpha(\mathbf
m(t))=1.$ Then, the transmission schedule is determined as follows. For any link $i$ in $\mathbf m(t)$, if no
links in $\mathcal{C}(i)$ were active in the previous data slot, then link $i$ is chosen to be \emph{active}
with an \emph{activation probability} $p_i$ and \emph{inactive} with probability $1-p_i$ in the current data
slot. If at least one link in $\mathcal{C}(i)$ was active in the previous data slot, then $i$ will be
inactive in the current data slot. Any link not selected by $\mathbf m(t)$ will maintain its state (active or
inactive) from the previous data slot. Conditions on the set of decision schedules $\mathcal M_0$ and the
link activation probabilities $p_i$'s will be specified later.\\

\noindent\hrulefill
\\
\textbf{Basic Scheduling Algorithm (in Time Slot $t$)}

\noindent\hrulefill
\begin{itemize}
\formula

\item[1.] In the control slot, randomly select a decision schedule $\mathbf m(t)\in \mathcal M_0$ with
probability $\alpha(\mathbf m(t))$.

\item[] $\forall i\in \mathbf m(t)$:\\
\phantom{aa} \textbf{If} no links in $\mathcal{C}(i)$ were active in the previous data slot, i.e., $\sum_{j\in \mathcal{C}(i)} x_j(t-1)=0$  \\
\phantom{aaaa} (a) $x_i(t)=1$ with probability $p_i$, $0<p_i<1$; \\
\phantom{aaaa} (b) $x_i(t)=0$ with probability $\bar{p}_i=1-p_i$. \\
\phantom{aa} \textbf{Else}\\
\phantom{aaaa} (c) $x_i(t)=0$.

\item[] $\forall i\notin \mathbf m(t):$\\
\phantom{aaaa} (d) $x_i(t)=x_i(t-1)$.

\item[2.] In the data slot, use $\mathbf x(t)$ as the transmission schedule.
\end{itemize}
\noindent\hrulefill \\

Note that the algorithm is a generalization of the so-called Glauber dynamics from statistical physics
\cite{mar99}, where multiple links are allowed to update their states in a single time slot. First we will
show that if the transmission schedule used in the previous data slot and the decision schedule selected in
the current control slot both are feasible, then the transmission schedule generated in the current data slot
is also feasible.

\begin{lemma}  \label{lemma:feasible}
\emph{If $\mathbf x(t-1) \in \mathcal{M}$ and $\mathbf m(t) \in \mathcal{M}$, then  $\mathbf x(t) \in
\mathcal{M}$.}
\end{lemma}
\begin{proof}
Note that $\mathbf x \in \mathcal{M}$ if and only if $\forall i\in E$ such that $x_i=1$, we have $x_j=0$ for
all $j\in \mathcal{C}(i)$.

Now consider any $i\in E$ such that $x_i(t)=1$. If $i \notin \mathbf m(t)$, then we know $x_i(t-1)=x_i(t)=1$,
and since $\mathbf x(t-1) \in \mathcal M$, we have $\forall j \in \mathcal{C}(i)$, $x_j(t-1)=0$. In addition,
if $j \notin \mathbf m(t)$, then $x_j(t)=x_j(t-1)=0$ based on Step (d) of the scheduling algorithm above;
otherwise, $j\in \mathbf m(t)$, then since $i \in \mathcal{C}(j)$ and $x_i(t-1)=1$, $x_j(t)=0$ based on Step
(c).

On the other hand, if $i \in \mathbf m(t)$, from the scheduling algorithm we have $x_i(t)=1$ only if
$x_j(t-1)=0$, $\forall j \in \mathcal{C}(i)$. Since $i \in \mathbf m(t)$ and $\mathbf m(t)$ is feasible, we
know $\mathcal{C}(i) \cap \mathbf m(t)=\emptyset$. Therefore, for any $j \in \mathcal{C}(i)$,
$x_j(t)=x_j(t-1)=0$ based on Step (d).
\end{proof}

Because $\mathbf x(t)$ only depends on the previous state $\mathbf x(t-1)$ and some randomly selected
decision schedule $\mathbf m(t)$, $\mathbf x(t)$ evolves as a discrete-time Markov chain (DTMC). Next we will
derive the transition probabilities between the states (transmission schedules).

\begin{lemma} \label{lemma:transition}
\emph{A state $\mathbf x \in \mathcal{M}$ can make a transition to a state $\mathbf x' \in \mathcal{M}$ if
and only if $\mathbf x \cup \mathbf x' \in \mathcal{M}$ and there exists a decision schedule $\mathbf m\in
\mathcal{M}_0$ such that
$$\mathbf x \bigtriangleup \mathbf x'=(\mathbf x \setminus \mathbf x')\cup(\mathbf x'
\setminus \mathbf x) \subseteq \mathbf m,$$
and in this case the transition probability from $\mathbf x$ to
$\mathbf x'$ is:}
\begin{eqnarray} \label{eq:transition}
P(\mathbf x, \mathbf x')  & = & \sum_{\mathbf m\in \mathcal M_0 : \mathbf x \bigtriangleup \mathbf x'
\subseteq \mathbf m}\alpha(\mathbf m) \Big(\prod_{l\in \mathbf x \setminus \mathbf x'}\bar{p}_l\Big)
\Big(\prod_{k\in \mathbf x'\setminus \mathbf x}p_k\Big)
 \nonumber \\
 & &  \Big(\prod_{i\in \mathbf m \cap (\mathbf x \cap \mathbf x')}p_i\Big) \Big(\prod_{j\in \mathbf m \setminus
(\mathbf x\cup \mathbf x') \setminus \mathcal{C}(\mathbf x \cup \mathbf x')}\bar{p}_j\Big).
\end{eqnarray}
\end{lemma}

\begin{proof}
(\emph{Necessity}) Suppose $\mathbf x$ is the current state and $\mathbf x'$ is the next state. $\mathbf x
\setminus \mathbf x'=\{l: \mathbf x_l=1, \mathbf x'_l=0\}$ is the set of links that change their state from 1
(active) to 0 (inactive). $\mathbf x' \setminus \mathbf x=\{k: \mathbf x_k=0, \mathbf x'_k=1\}$ is the set of
links that change their state from 0 to 1. From the scheduling algorithm, a link can change its state only if
the link belongs to the decision schedule. Therefore, $\mathbf x$ can make a transition to $\mathbf x'$ only
if there exists an $\mathbf m \in \mathcal{M}_0$ such that the symmetric difference $\mathbf x \bigtriangleup
\mathbf x' = (\mathbf x \setminus \mathbf x')\cup(\mathbf x' \setminus \mathbf x) \subseteq \mathbf m$. In
addition, since
$$(\mathbf x\cap \mathbf x')\cup(\mathbf x\setminus \mathbf x') = \mathbf x \in
\mathcal{M},$$
$$(\mathbf x\cap \mathbf x')\cup(\mathbf x'\setminus \mathbf x) = \mathbf x' \in \mathcal{M},$$
$$(\mathbf x\setminus \mathbf x')\cup(\mathbf x\setminus \mathbf x') = \mathbf x \bigtriangleup \mathbf x'\in
\mathcal{M},$$ we have
$$\mathbf x\cup \mathbf x' = (\mathbf x\setminus \mathbf x')\cup(\mathbf x\setminus
\mathbf x')\cup(\mathbf x\cap \mathbf x') \in \mathcal{M}.$$

(\emph{Sufficiency}) Now suppose $\mathbf x \cup \mathbf x' \in \mathcal{M}$ and there exists an $\mathbf m
\in \mathcal{M}_0$ such that $\mathbf x \bigtriangleup \mathbf x' \subseteq \mathbf m$. Given $\mathbf m$ is
the selected decision schedule, we can calculate the (conditional) probability that $\mathbf x$ makes a
transition to $\mathbf x'$, by dividing the links in $\mathbf m$ into the following five cases:
\begin{itemize}

\item[(1)] $l\in \mathbf x \setminus \mathbf x'$: link $l$ decides to change its state from 1 to 0, this
occurs with probability $\bar{p}_l$ based on Step (b) in the scheduling algorithm;

\item[(2)] $k\in \mathbf x' \setminus \mathbf x$: link $k$ decides to change its state from 0 to 1, this
occurs with probability $p_k$ based on Step (a);

\item[(3)] $i \in \mathbf m \cap (\mathbf x \cap \mathbf x') $: link $i$ decides to keep its state 1, this
occurs with probability $p_i$ based on Step (a);

\item[(4)] $e \in \mathbf m \cap \mathcal{C}(\mathbf x)$ where $\mathcal{C}(\mathbf x)=\cup_{l \in \mathbf
x}\mathcal{C}(l)$: link $e$ has to keep its state 0, this occurs with probability 1 based on Step (c);

\item[(5)] $j \in \mathbf m \setminus (\mathbf x\cup \mathbf x') \setminus \mathcal{C}(\mathbf x)$: link $j$
decides to keep its state 0, this occurs with probability $\bar{p}_j$ based on Step (b).
\end{itemize}

Note that $\mathbf m \cap \mathcal{C}(\mathbf x'\setminus \mathbf x) = \emptyset$ because $\mathbf
x'\setminus \mathbf x \subseteq \mathbf m$, we have $\mathbf m \setminus (\mathbf x\cup \mathbf x') \setminus
\mathcal{C}(\mathbf x) = \mathbf m \setminus (\mathbf x \cup \mathbf x') \setminus \mathcal{C}(\mathbf x \cup
\mathbf x').$ Since each link in $\mathbf m$ makes its decision independently of each other, we can multiply
these probabilities together. Summing over all possible decision schedules, we get the total transition
probability from $\mathbf x$ to $\mathbf x'$ given in (\ref{eq:transition}).
\end{proof}

\begin{proposition} \label{prop:productform}
\emph{A necessary and sufficient condition for the DTMC of the transmission schedules to be irreducible and
aperiodic is
\begin{eqnarray}
\cup_{\mathbf m\in \mathcal M_0}\mathbf m = E,
\end{eqnarray}
and in this case the DTMC is reversible and has the following product-form stationary distribution:}
\begin{eqnarray}
\pi(\mathbf x) & = & \frac{1}{Z}\prod_{i \in \mathbf x}\frac{p_i}{\bar{p}_i}, \label{eq:productform} \\
Z & = & \sum_{\mathbf x \in \mathcal M} \prod_{i \in \mathbf x}\frac{p_i}{\bar{p}_i}.
\end{eqnarray}
\end{proposition}

\begin{proof}
If $\cup_{\mathbf m\in \mathcal M_0}\mathbf m \neq E$, suppose $l \notin \cup_{\mathbf m\in \mathcal
M_0}\mathbf m$, then from state $\mathbf 0$ the DTMC will never reach a feasible schedule including $l$.
(There exists at least one such schedule, e.g., the schedule with only $l$ being active.)

On the other hand if $\cup_{\mathbf m\in \mathcal M_0}\mathbf m = E$, then using Lemma~\ref{lemma:transition}
it is easy to verify that state $\mathbf 0$ can reach any other state $\mathbf x \in \mathcal M$ with
positive probability in a finite number of steps and vice versa. To prove this, suppose $\mathbf
x=\{l_1,l_2,...,l_m\}$. Define $\mathbf x_j=\{l_1,...,l_j\}$ for $j=0,...,m$. Note that $\mathbf x_0= \mathbf
0$ and $\mathbf x_m =\mathbf x$. Now for $0 \leq j \leq m-1$, $\mathbf x_j \cup \mathbf x_{j+1}=\mathbf
x_{j+1}\in \mathcal{M}$ and $\mathbf x_j \bigtriangleup \mathbf x_{j+1}=\{l_{j+1}\}.$ Since $\cup_{\mathbf
m\in \mathcal M_0}\mathbf m = E$, there exists an $\mathbf m\in \mathcal{M}_0$ such that $\{l_{j+1}\}
\subseteq \mathbf m$. Then by Lemma~\ref{lemma:transition}, $\mathbf x_j$ can make a transition to $\mathbf
x_{j+1}$ with positive probability as given in (\ref{eq:transition}), hence $\mathbf 0$ can reach $\mathbf x$
with positive probability in a finite number of steps. The reverse argument is similar. Therefore, the DTMC
is irreducible and aperiodic.

In addition, if state $\mathbf x $ can make a transition to state $\mathbf x'$, then we can check that the
distribution in (\ref{eq:productform}) satisfies the detailed balance equation:
\begin{eqnarray}
\pi(\mathbf x)P(\mathbf x, \mathbf x') & = & \pi(\mathbf x')P(\mathbf x', \mathbf x),
\end{eqnarray}
hence the DTMC is reversible and (\ref{eq:productform}) is indeed its stationary distribution (see, for
example, \cite{kel79}).
\end{proof}

\subsection{Comments On Throughput-Optimality} \label{sec:throughputoptimal}

Based on the product-form distribution in Proposition~\ref{prop:productform}, and by choosing the link
activation probabilities as appropriate functions of the queue lengths, one can then proceed as in
\cite{jiawal08} (under a time-scale separation assumption) or as in \cite{rajshashi08} (without such an
assumption) to establish throughput-optimality of the scheduling algorithm. Instead of pursuing such a proof
here, we point out an alternative simple proof of throughput-optimality under the time-scale separation
assumption in \cite{jiawal08}.

We associate each link $i \in E$ with a nonnegative weight $w_i(t)$ in time slot $t$. Recall that MWS selects
a maximum-weight schedule $\mathbf x^*(t)$ in every time slot $t$ such that
\begin{eqnarray}
\sum_{i\in \mathbf x^*(t)}w_i(t) & = & \max_{\mathbf x \in \mathcal{M}} \sum_{i\in \mathbf x}w_i(t).
\end{eqnarray}

Let $q_i(t)$ be the queue length of link $i$ at the beginning of time slot $t$. It was proved in
\cite{taseph92} that MWS is throughput-optimal if we let $w_i(t)=q_i(t)$. This result was generalized in
\cite{erysriper05} as follows. For all $i\in E$, let link weight $w_i(t)=f_i(q_i(t))$, where $f_i: [0,\infty]
\rightarrow [0,\infty]$ are functions that satisfy the following conditions:
\begin{itemize}
\item[(1)] $f_i(q_i)$ is a nondecreasing, continuous function with $\lim_{q_i\rightarrow
\infty}f_i(q_i)=\infty$.

\item[(2)] Given any $M_1>0$, $M_2>0$ and $0< \epsilon <1$, there exists a $Q<\infty$, such that for all $q_i
> Q$ and $\forall i$, we have
\begin{equation*} \label{eq:fcondition}
(1-\epsilon)f_i(q_i) \leq f_i(q_i-M_1) \leq f_i(q_i+M_2) \leq (1+\epsilon)f_i(q_i).
\end{equation*}
\end{itemize}

The following result was established in \cite{erysriper05}.
\begin{theorem} \label{theorem:stable}
\emph{For a scheduling algorithm, if given any $\epsilon$ and $\delta$, $0 < \epsilon, \delta <1$, there
exists a $B>0$ such that: in any time slot $t$, with probability greater than $1-\delta$, the scheduling
algorithm chooses a schedule $\mathbf x(t)\in \mathcal M$ that satisfies}
\begin{eqnarray}
\sum_{i\in \mathbf x(t)}w_i(t) & \geq & (1-\epsilon) \max_{\mathbf x\in \mathcal M}\sum_{i\in \mathbf
x}w_i(t)
\end{eqnarray}
\emph{whenever $||\mathbf q(t)|| > B$, where $\mathbf q(t)=(q_i(t): i\in E)$. Then the scheduling algorithm
is throughput-optimal.}
\end{theorem}

If we choose the link activation probability
\begin{eqnarray}
p_i=\frac{e^{w_i(t)}}{e^{w_i(t)}+1}, \mbox{ } \forall i \in E,
\end{eqnarray}
then (\ref{eq:productform}) becomes
\begin{eqnarray}  \label{eq:productformweight}
\pi(\mathbf x) & = & \frac{1}{Z}\prod_{i \in \mathbf x}\frac{p_i}{\bar{p}_i} = \frac{1}{Z}\prod_{i \in
\mathbf x}e^{w_i(t)} \nonumber \\
& = & \frac{e^{\sum_{i\in \mathbf x}w_i(t)}}{Z}.
\end{eqnarray}
Hence the (steady-state) probability of choosing a schedule is proportional to its weight, so the schedules
with large weight will be selected with high probability. This is the intuition behind our proof.

By appropriately choosing the link weight functions $f_i$'s, we can make the DTMC of the transmission
schedules converge much faster compared to the dynamics of the link weights. For example, $f_i(q_i)=\alpha
q_i$ with a small $\alpha$ is suggested as a heuristic to satisfy the time-scale separation assumption in
\cite{jiawal08} and $f_i(q_i)=\log\log (q_i+e)$ is used in the proof of throughput-optimality in
\cite{rajshashi08} to essentially separate the time scales. Here, as in \cite{jiawal08}, we simply assume
that the DTMC is in the steady-state in every time slot.

\begin{proposition}  \label{prop:basicoptimal}
\emph{Suppose the basic scheduling algorithm satisfies $\cup_{\mathbf m\in \mathcal M_0}\mathbf m = E$ and
hence has the product-form stationary distribution. Let $p_i=\frac{e^{w_i(t)}}{e^{w_i(t)}+1}$, $\forall i \in
E$, where $w_i(t)$s are appropriate functions of the queue lengths. Then the scheduling algorithm is
throughput-optimal.}
\end{proposition}

\begin{proof}
We prove the proposition using Theorem~\ref{theorem:stable}. Now given any $\epsilon$ and $\delta$ such that
$0 < \epsilon, \delta <1$. Let $w^*(t):=\max_{\mathbf x\in \mathcal M}\sum_{i\in \mathbf x}w_i(t).$ Define
\begin{equation*}
\mathcal{X} := \Big\{\mathbf x\in \mathcal M: \mbox{ }\sum_{i\in \mathbf x}w_i(t) < (1-\epsilon)w^*(t)
\Big\}.
\end{equation*}

Since the DTMC has the product-form stationary distribution in (\ref{eq:productformweight}), we have
\begin{eqnarray}
\pi(\mathcal X) & = & \sum_{\mathbf x\in \mathcal X}\pi(\mathbf x) = \sum_{\mathbf x\in \mathcal X}\frac{e^{\sum_{i\in \mathbf x}w_i(t)}}{Z} \nonumber \\
& \leq & \frac{|\mathcal{X}|e^{(1-\epsilon)w^*(t)}}{Z} < \frac{2^{|E|}}{e^{\epsilon w^*(t)}},
\label{eq:bound}
\end{eqnarray}
where (\ref{eq:bound}) is true because $|\mathcal{X}|\leq |\mathcal{M}|\leq 2^{|E|}$, and
$$Z>e^{\max_{\mathbf x\in \mathcal M}\sum_{i\in \mathbf x}w_i(t)}=e^{w^*(t)}.$$

Therefore, if
\begin{eqnarray} \label{eq:condition}
w^*(t) & > & \frac{1}{\epsilon}\Big(|E|\log2+\log\frac{1}{\delta}\Big),
\end{eqnarray}
then $\pi(\mathcal{X})<\delta$. Since $w^*(t)$ is a continuous, nondecreasing function of $q_i(t)$'s, with
$\lim_{||\mathbf q(t)||\rightarrow \infty}w^*(t)=\infty$, there exits a $B>0$ such that whenever $||\mathbf
q(t)||>B$, (\ref{eq:condition}) holds and then $\pi(\mathcal{X})<\delta$. Hence the scheduling algorithm
satisfies the condition of Theorem~\ref{theorem:stable} and is throughput-optimal.
\end{proof}

\section{Distributed Implementation: Q-CSMA} \label{sec:distributed}

In this section we present a distributed implementation of the basic scheduling algorithm. The key idea is to
develop a distributed randomized procedure to select a (feasible) decision schedule in the control slot. To
achieve this, we further divide the control slot into control mini-slots. Note that once a link knows whether
it is included in the decision schedule, it can determine its state in the data slot based on its carrier
sensing information (i.e., whether its conflicting links were active in the previous data slot) and
activation probability. We call this implementation \textbf{Q-CSMA} (\textbf{Queue-length based CSMA/CA}),
since the activation probability of a link is determined by its queue length to achieve maximum throughput
(as in Section~\ref{sec:throughputoptimal}), and collisions of data packets are avoided via carrier sensing
and the exchange of control messages.

At the beginning of each time slot, every link $i$ will select a random backoff time. Link $i$ will send a
message announcing its INTENT to make a decision at the expiry of this backoff time subject to the
constraints described below.\\

\noindent\hrulefill
\\
\textbf{Q-CSMA Algorithm (at Link $i$ in Time Slot $t$)}

\noindent\hrulefill
\begin{itemize}
\formula

\item[1.] Link $i$ selects a random (integer) backoff time $T_i$ uniformly in $[0,W-1]$ and waits for $T_i$
control mini-slots. %(The idea is that $i$ will send a message announcing its INTENT to make a decision at the
%expiry of this backoff time subject to the constraints described in the next steps.)

\item[2.] IF link $i$ hears an INTENT message from a link in $\mathcal{C}(i)$ before the $(T_i+1)$-th control
mini-slot, $i$ will not be included in $\mathbf m(t)$ and will not transmit an INTENT message anymore. Link
$i$ will set $x_i(t)=x_i(t-1)$.
%\footnote{For example, the INTENT message can be an RTD/CTD pair
%exchanged by the sender and receiver of a link, and we select the duration of a control mini-slot such that
%every link can hear the INTENT message from its interfering links within one mini-slot.}

\item[3] IF link $i$ does not hear an INTENT message from any link in $\mathcal{C}(i)$ before the
$(T_i+1)$-th control mini-slot, it will send (broadcast) an INTENT message to all links in $\mathcal{C}(i)$
at the beginning of the $(T_i+1)$-th control mini-slot.

\begin{itemize}
\item If there is a collision (i.e., if there is another link in $\mathcal{C}(i)$ transmitting an INTENT
message in the same mini-slot), link $i$ will not be included in $\mathbf m(t)$ and will set
$x_i(t)=x_i(t-1)$.

\item If there is no collision, link $i$ will be included in $\mathbf m(t)$ and decide its state as follows: \\
\phantom{aa} \textbf{if} no links in $\mathcal{C}(i)$ were active in the previous data slot\\
\phantom{aaaaa} $x_i(t)=1$ with probability $p_i$, $0<p_i<1$; \\
\phantom{aaaaa} $x_i(t)=0$ with probability $\bar{p}_i=1-p_i$. \\
\phantom{aa} \textbf{else}\\
\phantom{aaaaa} $x_i(t)=0$.
\end{itemize}

\item[4.] IF $x_i(t)=1$, link $i$ will transmit a packet in the data slot.
%          If $x_i(t)=0$, $i$ will conduct carrier sensing in the data slot and set $NA_i(t)=1$ if it senses a data transmission and
%          $NA_i(t) = 0$ otherwise.
\end{itemize}
\noindent\hrulefill
\\

\begin{lemma}  \label{lemma:windowsize}
\emph{$\mathbf m(t)$ produced by Q-CSMA is a feasible schedule. Let $\mathcal M_0$ be the set of all decision
schedules produced by Q-CSMA. If the window size $W\geq 2$, then $\cup_{\mathbf m\in \mathcal M_0}\mathbf m =
E$.}
\end{lemma}

\begin{proof}
Under Q-CSMA, link $i$ will be included in the decision schedule $\mathbf m(t)$ if and only if it
successfully sends an INTENT message to all links in $\mathcal{C}(i)$ without a collision in the control
slot. This will ``silence'' the links in $\mathcal{C}(i)$ so those links will not be included in $\mathbf
m(t)$. Hence $\mathbf m(t)$ is feasible.

Now for any maximal schedule $\mathbf m$ (a schedule is \emph{maximal} if no additional links can be added to
the schedule without violating its feasibility), note that $\mathbf m$ will be selected in the control slot
if $T_i=0$, $\forall i \in \mathbf m$, and $T_j=1$, $\forall j \notin \mathbf m$. This occurs with positive
probability if $W\geq 2$, because,
\begin{eqnarray*}
\alpha(\mathbf m)  \geq  \mbox{Pr}\big\{T_i=0, \forall i \in \mathbf m; \mbox{ }T_j=1, \forall j \notin
\mathbf m\big\} = \prod_{i\in E}\frac{1}{W}>0.
\end{eqnarray*}

Since the set of all maximal schedules will include all links, $\cup_{\mathbf m\in \mathcal M_0}\mathbf m =
E$ if $W\geq 2$.
\end{proof}

Combining Lemma~\ref{lemma:windowsize} and Propositions~\ref{prop:productform},~\ref{prop:basicoptimal}  we
have the following result.
\begin{proposition}\label{prop:main-result}
\emph{Q-CSMA has the product-form distribution given in Proposition~\ref{prop:productform} if $W\geq 2$.
Further, it is throughput-optimal if we let $p_i=\frac{e^{w_i(t)}}{e^{w_i(t)}+1}$, $\forall i \in E$, where
$w_i(t)$s are appropriate functions of the queue lengths.}
\end{proposition}

\begin{remark} \label{remark:overhead}
A control slot of Q-CSMA consists of $W$ mini-slots and each link needs to send at most one INTENT message.
Hence Q-CSMA has constant (and low) signalling/time overhead, independent of the size of the network. Suppose
the duration of a data slot is $D$ mini-slots. Taking control overhead into account, Q-CSMA can achieve
$\frac{D}{D+W}$ of the capacity region, which approaches the full capacity when $W \ll D$.
\end{remark}

\begin{remark}
We can slightly modify Q-CSMA as follows: in Step 3, if link $i$ does not hear an INTENT message from any
link in $\mathcal{C}(i)$ before the $(T_i+1)$-th control mini-slot, $i$ will send an INTENT message to all
links in $\mathcal{C}(i)$ at the beginning of the $(T_i+1)$-th control mini-slot with some (positive)
probability $a_i$. In this case we can show that Q-CSMA achieves the product-form distribution even for
$W=1$. (We thank Libin Jiang for this observation.)
\end{remark}

When describing the Q-CSMA algorithm, we treat every link as an entity, while in reality each link consists
of a sender node and a receiver node. Both carrier sensing and transmission of data/control packets are
actually conducted by the nodes. In Appendix~\ref{appen:nodeQCSMA} we provide details to implement Q-CSMA
based on the nodes in the network. Such an implementation also allows us to resolve the hidden and exposed
terminal problems associated with wireless networks \cite{MACAW}, see Appendix~\ref{appen:hidden_exposed}.

\section{A Low-Delay Hybrid Q-CSMA Algorithm} \label{sec:hybrid}

By Little's law, the long-term average queueing delay experienced by the packets is proportional to the
long-term average queue length in the network. In our simulations (see Section \ref{sec:simulation}) we find
that the delay performance of Q-CSMA can be quite bad and much worse than greedy maximal scheduling GMS (this
is also true in simulations of the continuous-time CSMA algorithm). However, GMS is a centralized algorithm
and is not throughput-optimal in general (there exist networks, e.g., the $9$-link ring network in
Section~\ref{sec: ring_net}, where GMS can only achieve $2/3$ of the capacity region).

We are therefore motivated to design a distributed scheduling algorithm that can combine the advantages of
both Q-CSMA (for achieving maximum throughput) and GMS (for achieving low delay).  We first develop a
distributed algorithm to approximate GMS, which we call D-GMS.

The basic idea of D-GMS is to assign smaller backoff times to links with larger queue lengths. However, to
handle cases where two or more links in a neighborhood have the same queue length, some collision resolution
mechanism is incorporated in D-GMS. Further, we have conducted extensive simulations to understand how to
reduce the control overhead required to implement D-GMS while maintaining the ability to control the network
when the queue lengths become large. Based on these simulations, we conclude that it is better to use the log
of the queue lengths (rather than the queue lengths themselves) to determine the channel access priority of
the links. The resulting D-GMS algorithm is described below.

\vspace{0.1in} \noindent\hrulefill
\\
\textbf{D-GMS Algorithm (at Link $i$ in Time Slot $t$)}

\noindent\hrulefill
\begin{itemize}
\formula

\item[1.] Link $i$ selects a random backoff time
\begin{equation*}
T_i=W\times \lfloor B-\log_b\big(q_i(t)+1\big) \rfloor^{+}+\mbox{Uniform}[0,W-1]
\end{equation*}
and waits for $T_i$ control mini-slots.

\item[2.] IF link $i$ hears an RESV message (e.g., an RTS/CTS pair) from a link in $\mathcal{C}(i)$ before
the $(T_i+1)$-th control mini-slot, it will not be included in $\mathbf x(t)$ and will not transmit an RESV
message. Link $i$ will set $x_i(t)=0$.

\item[3.] IF link $i$ does not hear an RESV message from any link in $\mathcal{C}(i)$ before the $(T_i+1)$-th
control mini-slot, it will send an RESV message to all links in $\mathcal{C}(i)$ at the beginning of the
$(T_i+1)$-th control mini-slot.
\begin{itemize}
\item If there is a collision, link $i$ will set $x_i(t)=0$.

\item If there is no collision, link $i$ will set $x_i(t)=1$.
\end{itemize}
\item[4.] IF $x_i(t)=1$, link $i$ will transmit a packet in the data slot.
\end{itemize}
\noindent\hrulefill \vspace{0.1in}
\begin{remark}
In the above algorithm, each control slot can be thought as $B$ frames, with each frame consisting of $W$
mini-slots. Links are assigned a frame based on the log of their queue lengths and the $W$ mini-slots within
a frame are used to resolve contentions among links. Hence a control slot of D-GMS consists of $W\times B$
mini-slots, and links with empty queues will not compete for the channel in this time slot.
\end{remark}
\

Now we are ready to present a \emph{hybrid} Q-CSMA algorithm which is both provably throughput-optimal and
has very good delay performance in simulations. The basic idea behind the algorithm is as follows. For links
with weight greater than a threshold $w_0$, the Q-CSMA procedure (as in Section \ref{sec:distributed}) is
applied first to determine their states; for other links, the D-GMS procedure is applied next to determine
their states. To achieve this, a control slot is divided into $W_0$ mini-slots which are used to perform
Q-CSMA for links whose weight is greater than $w_0$ and $W_1\times B$ mini-slots which are used to implement
D-GMS among the other links. Each link $i$ uses a one-bit memory $N\!A_i$ to record whether any of its
conflicting links becomes active due to the Q-CSMA procedure in a time slot. This information is used in
constructing a schedule in the next
time slot.\\

\noindent\hrulefill
\\
\textbf{Hybrid Q-CSMA (at Link $i$ in Time Slot $t$)}

\noindent\hrulefill
%{\formula Link $i$ sets $N\!A_i(t)=0$.}
\begin{itemize}
\formula

\item[\textbf{IF}] $w_i(t)>w_0$ (\textbf{Q-CSMA Procedure})

\item[1.1] Link $i$ selects a random backoff time
$$T_i=\mbox{Uniform}[0,W_0-1].$$

\item[1.2] If link $i$ hears an INTENT message from a link in $\mathcal{C}(i)$ before the $(T_i+1)$-th
control mini-slot, then it will set $x_i(t)=x_i(t-1)$ and go to Step 1.4.

\item[1.3] If link $i$ does not hear an INTENT message from any link in $\mathcal{C}(i)$ before the
$(T_i+1)$-th control mini-slot, it will send an INTENT message to all links in $\mathcal{C}(i)$ at the
beginning of the $(T_i+1)$-th control mini-slot.

\item If there is a collision, link $i$ will set $x_i(t)=x_i(t-1)$.

\item If there is no collision, link $i$ will decide its state as follows: \\
\phantom{aaa} \textbf{if} no links in $\mathcal{C}(i)$ were active due to the Q-CSMA procedure in the previous data slot, i.e., $N\!A_i=0$\\
\phantom{aaaaaa} $x_i(t)=1$ with probability $p_i$, $0<p_i<1$; \\
\phantom{aaaaaa} $x_i(t)=0$ with probability $\bar{p}_i=1-p_i$. \\
\phantom{aaa} \textbf{else}\\
\phantom{aaaaaa} $x_i(t)=0$.

\item[1.4] If $x_i(t)=1$, link $i$ will send an RESV message to all links in $\mathcal{C}(i)$ at the
beginning of the $(W_0+1)$-th control mini-slot. It will set $N\!A_i=0$ and transmit a packet in the data slot.\\
           If $x_i(t)=0$ and link $i$ hears an RESV message from any link in $\mathcal{C}(i)$ in the $(W_0+1)$-th control mini-slot, it will set
           $N\!A_i=1$; otherwise, it will set $N\!A_i=0$.

%$i$ will transmit from the $W_0$-th control mini-slot through the
%data slot. Otherwise, $i$ will keep silent. (Recall that we have
%assumed the service rate of each link to be $1$ packet/slot and here
%we further assume the packet length to be a fixed number of bits.
%Then the data transmission will actually terminate $B\times W_1$
%control mini-slots before the end of the whole time slot.)

\

\item[\textbf{IF}] $w_i(t)\leq w_0$ (\textbf{D-GMS Procedure})

\item[2.1] If link $i$ hears an RESV message from any link in $\mathcal{C}(i)$ in the $(W_0+1)$-th control
mini-slot, it will set $N\!A_i=1$ and $x_i(t)=0$ and keep silent in this time slot. \\
           Otherwise, link $i$ will set $N\!A_i=0$ and select a random backoff time
           \begin{eqnarray*}
           T_i & =& (W_0+1)+W_1\times \lfloor B-\log_b\big(q_i(t)+1\big) \rfloor^{+} \\
           & & +\mbox{Uniform}[0,W_1-1]
           \end{eqnarray*}
           and wait for $T_i$ control mini-slots.

\item[2.2] If link $i$ hears an RESV message from a link in $\mathcal{C}(i)$ before the $(T_i+1)$-th control
mini-slot, it will set $x_i(t)=0$ and keep silent in this time slot.

\item[2.3] If link $i$ does not hear an RESV message from any link in $\mathcal{C}(i)$ before the
$(T_i+1)$-th control mini-slot, it will send an RESV message to all links in $\mathcal{C}(i)$ at the
beginning of the $(T_i+1)$-th control mini-slot.

\begin{itemize}
\item If there is a collision, link $i$ will set $x_i(t)=0$.

\item If there is no collision, link $i$ will set $x_i(t)=1$.
\end{itemize}

\item[2.4] If $x_i(t)=1$, link $i$ will transmit a packet in the data slot.
\end{itemize}

\noindent\hrulefill \vspace{0.1in}
\begin{remark}
The $(W_0+1)$-th control mini-slot (called \emph{transition mini-slot}, which occurs between the first $W_0$
mini-slots and the last $W_1\times B$ mini-slots) is reserved for all the links which have not been scheduled
so far to conduct carrier sensing. In this mini-slot those links which have already been scheduled (due to
the Q-CSMA procedure) will send an RESV message so their neighbors can sense and record this information in
their $N\!A$ bit.
\end{remark}

\begin{remark}
Suppose that the link weights are chosen as in Section~\ref{sec:throughputoptimal}, i.e.,  $w_i(t) =
f_i(q_i(t))$ is an increasing function of $q_i(t)$. Thus, $w_i(t)\gtrless w_0$ is equivalent to
$q_i(t)\gtrless q_0$, where $q_0 = f^{-1}(w_0)$ is the \emph{queue-length threshold}.
\end{remark}

\begin{remark}
The control overhead of the hybrid Q-CSMA algorithm is $W_0+1+W_1\times B$ per time slot. As in the pure
D-GMS algorithm, links with empty queues will keep silent throughout the time slot.
\end{remark}

\subsection{Throughput-Optimality of Hybrid Q-CSMA Algorithm}

Let $L=\{i\in E: w_i(t)>w_0\}$ be the set of links for which the Q-CSMA procedure is applied (in time slot
$t$), and $L^c=E\setminus L$. Let $\mathbf x_L(t)=(x_i(t): i\in L)$ be the transmission schedule of the links
in $L$. Note that in the hybrid Q-CSMA algorithm, scheduling links in $L^c$ will not affect the Q-CSMA
procedure because those links will be scheduled after the links in $L$ and their transmissions will not be
recorded by their neighboring links in the $N\!A$ bits. Therefore, under fixed link weights and activation
probabilities (so $L$ is also fixed), $\mathbf x_L(t)$ evolves as a DTMC. Further, using similar arguments as
in the proofs of Propositions~\ref{prop:productform} and \ref{prop:main-result}, we have
\begin{proposition}
\emph{If $W_0\geq 2$, then the DTMC describing the evolution of the transmission schedule $\mathbf x_L(t)$ is
reversible and has the following product-form stationary distribution:}
\begin{eqnarray}
\pi(\mathbf x_L) & = & \frac{1}{Z_L}\prod_{i \in \mathbf x_L}\frac{p_i}{\bar{p}_i}, \label{eq:productform_L} \\
Z_L & = & \sum_{\mathbf x_L \in \mathcal M_L} \prod_{i \in \mathbf x_L}\frac{p_i}{\bar{p}_i},
\end{eqnarray}
\emph{where $\mathcal M_L$ is the set of feasible schedules when restricted to links in $L$.}
\end{proposition}

Assuming a time-scale separation property that the DTMC of $\mathbf x_L(t)$ is in steady-state in every time
slot, we establish the throughput-optimality of the hybrid Q-CSMA algorithm in the following proposition.

\begin{proposition} \label{prop:hybridoptimal}
\emph{For each link $i\in L$, we choose its activation probability $p_i=\frac{e^{w_i(t)}}{e^{w_i(t)}+1}$,
where the link weights $w_i(t)$'s are appropriate functions of the queue lengths as in
Section~\ref{sec:throughputoptimal}. Then the hybrid Q-CSMA algorithm is throughput-optimal.}
\end{proposition}

\begin{proof}
Write $\mathbf x(t)=(\mathbf x_L(t), \mathbf x_{L^c}(t))$, where $\mathbf x_A(t)=(x_i(t): i\in A)$ for any
set $A \subseteq E$. Recall that MWS selects a maximum-weight schedule $\mathbf x^*(t)$ such that
\begin{eqnarray*}
w(\mathbf x^*(t)) & = & w(\mathbf x^*_L(t)) + w(\mathbf x^*_{L^c}(t)) \\
& = & \max_{\mathbf x\in \mathcal M}\sum_{i\in \mathbf x}w_i(t) =: w^*,
\end{eqnarray*}
where $w(\mathbf x_A(t))=\sum_{i\in \mathbf x_A(t)} w_i(t)$.

It is clear that
\begin{equation}
w(\mathbf x^*_L(t)) \leq \max_{\mathbf x_L\in \mathcal M_L}w(\mathbf x_L)  =: w^*_L.
\end{equation}

For any $\epsilon$ such that $0<\epsilon<1$, when $||\mathbf w(t)||_\infty \geq \frac{2|E|w_0}{\epsilon}$ (so
$L$ is not empty), we have
\begin{equation*}
w^*_L \geq \max_i w_i(t) \geq \frac{2|E|w_0}{\epsilon}.
\end{equation*}

Therefore,
\begin{equation} \label{eq:weightbound}
w(\mathbf x^*_{L^c}(t)) \leq |E|  w_0 \leq \frac{\epsilon}{2}w_L^*.
\end{equation}

Using similar arguments as in the proof of Proposition~\ref{prop:basicoptimal}, we can show that for any
$\epsilon$ and $\delta$ such that $0 < \epsilon, \delta < 1$, if the queue lengths are large enough, then
with probability greater than $1-\delta$, the Q-CSMA procedure chooses $\mathbf x_L(t)$ such that
\begin{eqnarray*}
w(\mathbf x_L(t)) \geq (1-\frac{\epsilon}{2}) \max_{\mathbf x_L\in \mathcal M_L} w(\mathbf x_L) =
(1-\frac{\epsilon}{2}) w^*_L.
\end{eqnarray*}
In addition, if the queue lengths are large enough, then (\ref{eq:weightbound}) holds. Therefore, since
$w(\mathbf x_{L^c}(t)) \geq 0$, we have
\begin{eqnarray*}
w(\mathbf x(t)) & = & w(\mathbf x_L(t)) + w(\mathbf x_{L^c}(t)) \\
& \geq & (1-\epsilon) w^*_L + \frac{\epsilon}{2} w^*_L \nonumber \\
                & \geq & (1-\epsilon) w(\mathbf x^*_L(t)) + w(\mathbf x^*_{L^c}(t)) \\
                & \geq & (1-\epsilon) w^*.
\end{eqnarray*}
Hence the hybrid Q-CSMA algorithm satisfies the condition of Theorem~\ref{theorem:stable} and is
throughput-optimal.
\end{proof}

\begin{remark}
In the above algorithm, one can replace D-GMS by any other heuristic and still maintain
throughput-optimality. We simply use D-GMS because it is an approximation to GMS which is known to perform
well in a variety of previous simulation studies. It is also important to recall our earlier observation that
GMS is not a distributed algorithm and hence, we have to resort to a distributed approximation.
\end{remark}

\section{Simulation Results}   \label{sec:simulation}

In this section we evaluate the performance of different scheduling algorithms via simulations, which include
MWS (only for small networks), GMS (centralized), D-GMS, Q-CSMA, and the hybrid Q-CSMA algorithm. In
addition, we have implemented a distributed algorithm to approximate maximal scheduling (called D-MS), which
can be viewed as a synchronized slotted version of the IEEE 802.11 DCF with the RTS/CTS mechanism. Note that
D-MS is a special case of D-GMS presented in Section~\ref{sec:hybrid} with $B=1$ so the backoff time of a
link does not depend on its queue length.\\

\noindent\hrulefill
\\
\textbf{D-MS (at Link $i$ in Time Slot $t$)}

\noindent\hrulefill
\begin{itemize}
\formula

\item[1.] Link $i$ selects a random backoff time
$$T_i=\mbox{Uniform}[0,W-1]$$
and waits for $T_i$ control mini-slots.

\item[2.] If link $i$ hears an RESV message from a link in $\mathcal{C}(i)$ before the $(T_i+1)$-th control
mini-slot, it will not be included in the transmission schedule $\mathbf x(t)$ and will not transmit an RESV
message. Link $i$ will set $x_i(t)=0$.

\item[3.] If link $i$ does not hear an RESV message from any link in $\mathcal{C}(i)$ before the $(T_i+1)$-th
control mini-slot, it will send an RESV message to all links in $\mathcal{C}(i)$ at the beginning of the
$(T_i+1)$-th control mini-slot.
\begin{itemize}
\item If there is a collision, link $i$ will set $x_i(t)=0$.

\item If there is no collision, link $i$ will set $x_i(t)=1$.
\end{itemize}
\item[4.] If $x_i(t)=1$, link $i$ will transmit a packet in the data slot.

\item[] (Links with empty queues will keep silent in this time slot.)
\end{itemize}
\noindent\hrulefill \vspace{0.1in}

\subsection{A 24-Link Grid Network}
\label{sec: grid_net}
\begin{figure}[t]
\centering
\includegraphics[width=4.5cm]{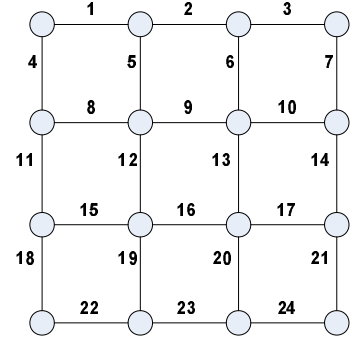}
\caption{A 24-link grid network topology.} \label{fig: grid_net}
\end{figure}

We first evaluate the performance of the scheduling algorithms in a grid network with 16 nodes and 24 links
as shown in Fig.~\ref{fig: grid_net}. Each node is represented by a circle and each link is illustrated by a
solid line with a label indicating its index. Each link maintains its own queue. We assume $1$-hop
interference.

Consider the following four sets of links:
$${\cal L}_1=\{1,3,8,10,15,17,22,24\},$$
$${\cal L}_2=\{4,5,6,7,18,19,20,21\},$$
$${\cal L}_3=\{1,3,9,11,14,16,22,24\},$$
$${\cal L}_4=\{2,4,7,12,13,18,21,23\}.$$
Each set represents a (maximum-size) maximal schedule of the network. Let $\mathbf{M}_i = \mathbf{e}_{{\cal
L}_i}$, where $\mathbf{e}_{{\cal L}_i}$ is a vector in which the components with indices in ${\cal L}_i$ are
$1$'s and others are $0$'s. Then, we let the arrival rate vector be a convex combination of those maximal
schedules scaled by $\rho$:
$$\bm{\lambda} = \rho \cdot \sum_{i=1}^4 c_i \mathbf{M}_i, ~\mathbf{c} =
[0.2,0.3,0.2,0.3].$$ Note that a convex combination of several maximum-size maximal schedules must lie on the
boundary of the capacity region. Hence the parameter $\rho$ in $[0,1]$ can be viewed as the \emph{traffic
intensity}, with $\rho \rightarrow 1$ representing arrival rates approaching the boundary of the capacity
region.

The packet arrivals to each link $i$ follow a Bernoulli process with rate $\lambda_i$ independent of the
packet arrival processes at other links. Each simulation experiment starts with all empty queues. For each
algorithm under a fixed $\rho$, we take the average over $10$ independent experiments, with each run being
$10^5$ time slots. Due to the high complexity of MWS in such a large network, we do not implement it here.

\begin{figure}[t]
\centering
\includegraphics[width=\columnwidth]{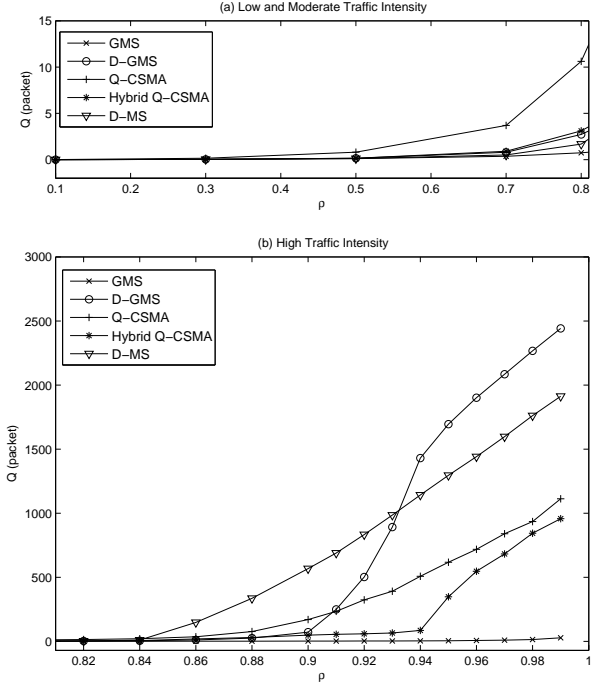}
\caption{Long-term average queue length per link in the 24-link grid network.} \label{fig: sim_grid}
\end{figure}

For fair comparison, we choose a control overhead of $48$ mini-slots for every distributed scheduling
algorithm (which lies in the range of the backoff window size specified in IEEE 802.11 DCF \cite{bia00}). The
parameter setting of the scheduling algorithms is summarized below.
\begin{itemize}
\item D-MS: $W = 48$.

\item D-GMS: $B = 3,~W = 16$; $b = 8$.

\item Q-CSMA: $W = 48$; link weight $w_i(t) = \log(0.1q_i(t))$ and link activation probability
$p_i=\frac{e^{w_i(t)}}{e^{w_i(t)}+1}$.

\item Hybrid Q-CSMA: $W_0=5$ for the Q-CSMA procedure, $B=3$ and $W_1=14$ for the D-GMS procedure, plus $1$
transition mini-slot. The queue-length threshold $q_0 = 100$. Other parameters are the same as Q-CSMA.
\end{itemize}

\begin{remark}
In Q-CSMA we choose the link weight function $w_i(t) = \log(\alpha q_i(t))$ with a small constant $\alpha$.
The rationality is to make the link weights change much slower than the dynamics of the CSMA Markov chain (to
satisfy the time-scale separation assumption). We have tried several other choices for the link weight
functions suggested in prior literature (such as $\alpha q_i(t)$ in \cite{jiawal08} and $\log\log(q_i(t)+e)$
in \cite{rajshashi08}) but $\log(\alpha q_i(t))$ seems to give the best delay performance.
\end{remark}

The performance of the scheduling algorithms is shown in Fig.~\ref{fig: sim_grid}, from which we can see
that:
\begin{itemize}
\item Under low to moderate traffic intensity, D-GMS and D-MS have very good delay performance (small
long-term average queue length) and perform better than Q-CSMA. However, when the traffic intensity is high,
the average queue length under D-GMS/D-MS blows up and their delay performance becomes much worse than
Q-CSMA.

\item Hybrid Q-CSMA has the best delay performance among the distributed scheduling algorithms. It retains
the stability property of Q-CSMA even under high traffic intensity while significantly reduces the delay of
pure Q-CSMA. Note that when $\rho \rightarrow 1$, the performance of Hybrid Q-CSMA becomes close to pure
Q-CSMA since the effect of the D-GMS procedure diminishes when the queue lengths of most links exceed the
queue length threshold.

\item Centralized GMS has excellent delay performance, but it is not throughput-optimal in general, as
illustrated next.
\end{itemize}

We have tested the algorithms under other traffic patterns, e.g., different arrival rate vectors, Poisson
arrivals, and the results are similar.

\subsection{A 9-Link Ring Network}
\label{sec: ring_net}

\begin{figure}[t]
\centering
\includegraphics[width=4cm]{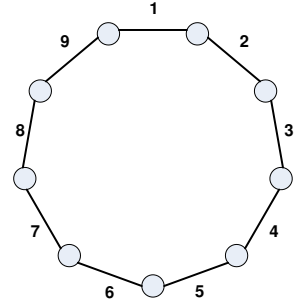}
\caption{A 9-link ring network topology.} \label{fig: ring_net}
\end{figure}

Consider a 9-link ring network under the $2$-hop interference model, as shown in Fig.~\ref{fig: ring_net}. It
was shown in \cite{lecnisri09} that GMS can only achieve $2/3$ of the capacity region in this network. To see
this, we construct a traffic pattern using the idea in \cite{joolinshr08}. Define
$${\cal L}_i = \{i,(i+4)\mod9\}, \mbox{ }1\le i \le 9.$$
Starting with empty queues, in time slot $9k+i~(k\in \mathbb{Z})$, one packet arrives at each of the $2$
links in ${\cal L}_i$, and, with probability $\epsilon$, an additional packet arrives at each of the $9$
links. The average arrival rate vector is then $\bm{\lambda} = (\frac{2}{9}+\epsilon) \mathbf{e}$, where
$\mathbf{e}$ is a vector with all components equal to $1$. It has been shown in \cite{joolinshr08} that GMS
will lead to infinite queue lengths under such a traffic pattern for all $\epsilon > 0$.

On the other hand, we could use a scheduling policy as follows. Define
$${\cal \tilde{L}}_1 = \{1,4,7\},~{\cal
\tilde{L}}_2 = \{2,5,8\},~{\cal \tilde{L}}_3 = \{3,6,9\},$$ and let $\mathbf{\tilde{M}}_i = \mathbf{e}_{{\cal
\tilde{L}}_i}$ for $1\le i \le 3.$ In time slot $3k+i~(k\in \mathbb{Z})$, the maximal schedule
$\mathbf{\tilde{M}}_i$ is used. Hence, the average service rate vector is $\bm{\mu} = \frac{1}{3}\mathbf{e}$.
When $0 < \epsilon < \frac{1}{9}$, $\bm{\lambda} < \bm{\mu}$, i.e., $\bm{\lambda}$ lies in the interior of
the capacity region, but GMS cannot keep the network stable as we saw above.

We evaluate the performance of the scheduling algorithms under the above traffic pattern. Each simulation
experiment starts with all empty queues. For each algorithm under a fixed $\epsilon$, we take the average
over $10$ independent experiments, with each run being $10^5$ time slots. We use exactly the same parameter
setting as in Section~\ref{sec: grid_net}.

\begin{figure}[t]
\centering
\includegraphics[width=\columnwidth]{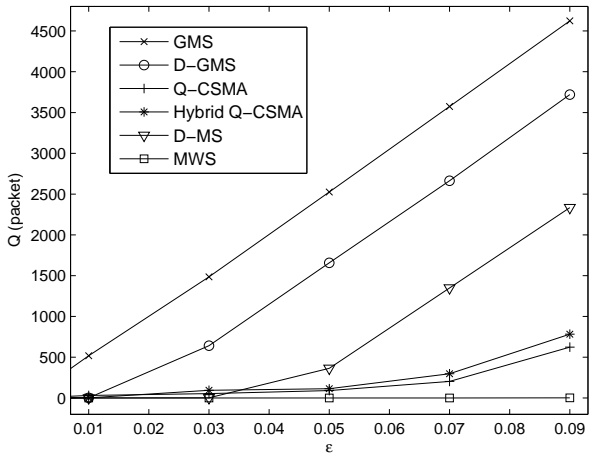}
\caption{Long-term average queue length per link in the 9-link ring network.} \label{fig: sim_ring}
\end{figure}

\begin{figure}[t]
\centering
\includegraphics[width=\columnwidth]{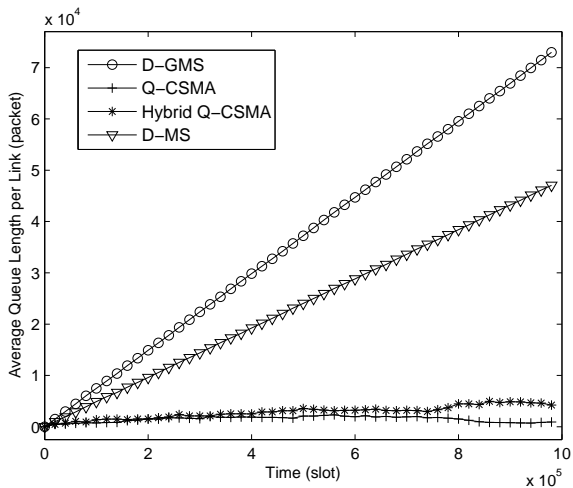}
\caption{Sample paths of average queue length per link: $\epsilon=0.09$.} \label{fig: sim_ring_sample_path}
\end{figure}

In Fig.~\ref{fig: sim_ring} we can see that Q-CSMA and Hybrid Q-CSMA have a much lower delay than GMS, D-GMS
(when $\epsilon \geq 0.03$) and D-MS (when $\epsilon \geq 0.05$). Fig.~\ref{fig: sim_ring_sample_path} shows
that the average queue length increases linearly with the running time ($\#$ of time slots) under D-GMS/D-MS
which imply that they are not stable, while the average queue length becomes stable under Q-CSMA/Hyrbid
Q-CSMA.

\section{Conclusion} \label{sec:conclusion}

In this paper, we proposed a discrete-time distributed queue-length based CSMA/CA protocol that leads to
collision-free data transmission schedules. The protocol is provably throughput-optimal. The discrete-time
formulation allows us to incorporate mechanisms to dramatically reduce the delay without affecting the
theoretical throughput-optimality property. In particular, combining CSMA with distributed GMS leads to very
good delay performance.

We believe that it should be straightforward to extend our algorithms to be applicable to networks with
multi-hop traffic and congestion-controlled sources (see \cite{linshrsri06,geoneetas06,shasri07} for related
surveys).

\appendix

\subsection{Node-Based Implementation of Q-CSMA} \label{appen:nodeQCSMA}

For any node $n$ in the network, we use $\mathcal{N}(n)$ to denote the neighborhood of $n$, which is the set
of nodes that can hear the transmission of $n$. We assume \emph{symmetry} in hearing: if $n'\in
\mathcal{N}(n)$ then $n\in \mathcal{N}(n')$.

Let $s_i$ and $r_i$ be the sender node and receiver node of link $i$, respectively. If link $i$ is included
in the transmission schedule, then in the data slot, $s_i$ will send a \textbf{data} packet to $r_i$, and
$r_i$ will reply an \textbf{ACK} packet to $s_i$. We assume that the data transmission from $s_i$ to $r_i$ is
successful if no nodes in $\mathcal{N}(r_i)$ are transmitting in the same time; similarly, the ACK
transmission from $r_i$ to $s_i$ is successful if no nodes in $\mathcal{N}(s_i)$ are transmitting in the same
time. We also consider the node-exclusive constraint that two active links cannot share a common node.
Therefore, in a \emph{synchronized} data/ACK transmission system, the conflict set of link $i$ is:
\begin{eqnarray*}
\mathcal{C}(i) & = & \Big\{ j: j \mbox{ shares a common node with }i, \\
& & \mbox{ or } s_j \in \mathcal{N}(r_i), \mbox{ or } r_j \in \mathcal{N}(s_i). \Big\}.
\end{eqnarray*}

In summary, two links $i$ and $j$ conflict with each other, i.e., $i\in \mathcal{C}(j)$ and $j\in
\mathcal{C}(i)$, if they share a common node, or if simultaneous data transmissions at $s_i$ and $s_j$ will
collide at $r_i$ and $r_j$, or if simultaneous ACK transmissions at $r_i$ and $r_j$ will collide at $s_i$ and
$s_j$.

We say that a node is \emph{active} in a time slot if it is the sender node or receiver node of an active
link. In a time slot, each \emph{inactive} node will conduct carrier sensing. It will determine whether there
are some active sender nodes and some active receiver nodes in its neighborhood by sensing whether the
channel is busy during the data transmission period and during the ACK transmission period, respectively. In
this way, link $i$ ``knows" that no links in its conflict set are active in a time slot, if $s_i$ and $r_i$
don't belong to an active link other than $i$, $s_i$ does not sense an active receiver node in
$\mathcal{N}(s_i)$, and $r_i$ does not sense an active sender node in $\mathcal{N}(r_i)$.

Similar to the RTS/CTS mechanism in the 802.11 MAC protocol, an INTENT message ``sent" by a link consists of
an RTD (\emph{Request-To-Decide}) and a CTD (\emph{Clear-To-Decide}) pair exchanged by the sender node and
receiver node of the link. To achieve this, we further divide a control mini-slot into two sub-mini-slots. In
the first sub-mini-slot, $s_i$ sends an RTD to $r_i$. If $r_i$ receives the RTD without a collision (i.e., no
nodes in $\mathcal{N}(r_i)$ are transmitting in the same sub-mini-slot), then $r_i$ will reply a CTD to $s_i$
in the second sub-mini-slot. If $s_i$ receives the CTD from $r_i$ without a collision, then link
$i=(s_i,r_i)$ will be added to the decision schedule. We choose the length of a sub-mini-slot such that an
RTD or CTD sent by any node can reach its neighbors within one sub-mini-slot. Note that the exchange of an
RTD/CTD pair between the sender and receiver of a link will ``silence" all its conflicting links so those
links will not be added to the decision schedule anymore.

Now we are ready to present the node-based Q-CSMA algorithm. Some additional one-bit memories maintained at
node $n$ (in time slot $t$) are summarized below (each explanation corresponds to bit $1$):
\begin{itemize}
\item $A\!S_n(t)/A\!R_n(t)$: $n$ is \emph{available} as the \emph{sender}/\emph{receiver} node for a link in
the decision schedule $\mathbf{m}(t)$.

\item $A\!C\!T_n(t)$: $n$ is \emph{active} (as either a sender or receiver node).

\item $N\!S_n(t)/N\!R_n(t)$: the \emph{neighborhood} of $n$ (i.e, ${\cal N}(n)$) has an active
\emph{sender}/\emph{receiver} node.
\end{itemize}

\noindent\hrulefill
\\
\textbf{Q-CSMA Algorithm (at Node $n$ in Time Slot $t$)}

\noindent\hrulefill

\begin{itemize}
\formula

\item[1.] At the beginning of the time slot, node $n$ sets $A\!S_n(t)=1$ and $A\!R_n(t)=1$. \\
Let $L(n)$ be the set of links for which $n$ is the sender node (i.e., $n=s_l$, $\forall l\in L(n)$). Node
$n$ randomly chooses one link in $L(n)$ (suppose link $i=(n,m)$ is chosen) and selects a backoff time $T_i$
uniformly in $[0,W-1]$. Other links in $L(n)$ will not be included in $\mathbf m(t)$, so $x_l(t)=x_l(t-1),
\forall l \in L(n)\setminus i$.

\item[2.] Throughout the control slot, if node $n$ senses an RTD transmission not intended for itself (or a
collision of RTDs) by a node in $\mathcal{N}(n)$, $n$ will no longer be available as the receiver node for a
link in $\mathbf m(t)$. Thus, node $n$ will set $A\!R_n(t)=0$.

\item[3.] Before the $(T_i+1)$-th control mini-slot, if node $n$ senses a CTD transmission by a node in
$\mathcal{N}(n)$ (or a collision of CTDs), $n$ will no longer be available as the sender node for a link in
$\mathbf m(t)$, and it will set $A\!S_n(t)=0$. In this case link $i$ will not be included in $\mathbf m(t)$
and $x_i(t)=x_i(t-1)$.

\item[4.] At the beginning of the $(T_i+1)$-th control mini-slot, if $A\!S_n(t)=1$, node $n$ will send an RTD
to node $m$ in the first sub-mini-slot. Node $n$ will then set $A\!S_n(t)=A\!R_n(t)=0$.

\begin{itemize}
\item[4.1] If node $m$ receives the RTD from node $n$ without a collision and $A\!R_m(t)=1$, $m$ will send a
CTD to $n$ in the second sub-mini-slot of the $(T_i+1)$-th control mini-slot. Node $m$ will then set
$A\!S_m(t)=A\!R_m(t)=0$. The CTD message also includes the carrier sensing information of node $m$ in the
previous time slot (the values of $A\!C\!T_m(t-1)$ and $N\!S_m(t-1)$). Otherwise, no message will be sent.

\item[4.2] If node $n$ receives the CTD message from node $m$ without a collision, link $i=(n,m)$ will be included in $\mathbf m(t)$. Node $n$ will decide link $i$'s state as follows: \\
\phantom{aaa} \textbf{if} no links in $\mathcal{C}(i)$ were active in the previous data slot, i.e., $x_i(t-1)=1$ or $A\!C\!T_n(t-1)=A\!C\!T_m(t-1)=N\!R_n(t-1)=N\!S_m(t-1)=0$\\
\phantom{aaaaaa} $x_i(t)=1$ with probability $p_i$, $0<p_i<1$; \\
\phantom{aaaaaa} $x_i(t)=0$ with probability $\bar{p}_i=1-p_i$. \\
\phantom{aaa} \textbf{else} \\
\phantom{aaaaaa} $x_i(t)=0$. \\
Otherwise, link $i$ will not be included in $\mathbf m(t)$ and $x_i(t)=x_i(t-1)$.
\end{itemize}

\item[5.] In the data slot, node $n$ takes one of the three different roles:
    \begin{itemize}

    \item \emph{Sender}: $x_l(t)=1$ for some link $l=(n,m)\in E$. Node $n$ will send a data packet to node $m$
    and set $A\!C\!T_n(t)=1$.

    \item \emph{Receiver}: $x_l(t)=1$ for some link $l=(m,n)\in E$. Node $n$ will send an ACK packet to node $m$
    (after it receives the data packet from $m$) and set $A\!C\!T_n(t)=1$.

    \item \emph{Inactive}: Node $n$ sets $A\!C\!T_n(t)=0$ and conducts carrier sensing. Recall that data/ACK
    transmissions in our system are synchronized. Thus, node $n$ will set $N\!S_n(t)=0$ if it senses no signal
    during the data transmission period and set $N\!S_n(t)=1$ otherwise. Similarly, node $n$ will set $N\!R_n(t)=0$ if it senses
    no signal during the ACK transmission period and set $N\!R_n(t)=1$ otherwise.
    \end{itemize}

\end{itemize}
\noindent\hrulefill \vspace{0.1in}

\begin{remark}
Note that RTD and CTD are sent in two different sub-mini-slots. This provides an easy way to differentiate
RTD and CTD (or collisions of RTDs and CTDs, respectively) without having to encode the packet type in a
preamble bit of such a control packet (actually when a collision happens, a node cannot even check this
``packet type'' bit to differentiate RTD and CTD).
\end{remark}

\

\begin{proposition}  \label{prop:nodeQCSMA}
\emph{$\mathbf m(t)$ produced by the node-based Q-CSMA algorithm is a feasible schedule. Let $\mathcal M_0$
be the set of decision schedules produced by the algorithm. If the window size $W\geq 2$, then $\cup_{\mathbf
m\in \mathcal M_0}\mathbf m = E$ and the algorithm achieves the product-form distribution in
Proposition~\ref{prop:productform}.}
\end{proposition}
\begin{proof}
The proof is similar to the proof of Lemma~\ref{lemma:windowsize}. Under the node-based Q-CSMA algorithm,
link $i$ will be included in the decision schedule $\mathbf m(t)$ if and only if its sender and receiver
nodes successfully exchange an RTD/CTD pair in the control slot. This will ``silence" all the receivers in
$\mathcal{N}(s_i)$ and all the senders in in $\mathcal{N}(r_i)$ as well as nodes $s_i$ and $r_i$, so no links
in $\mathcal{C}(i)$ will be included in $\mathbf m(t)$. Hence $\mathbf m(t)$ is a feasible schedule.
Similarly, for any maximal schedule $\mathbf m$, we can check that $\mathbf m$ will be selected in the
control slot with positive probability if the window size $W\geq 2$. Since the set of all maximal schedules
will include all links, $\cup_{\mathbf m\in \mathcal M_0}\mathbf m = E$. Then by
Proposition~\ref{prop:productform} the node-based Q-CSMA algorithm achieves the product-form distribution.
\end{proof}

\begin{figure}[t]
\centering
\includegraphics[width=6cm]{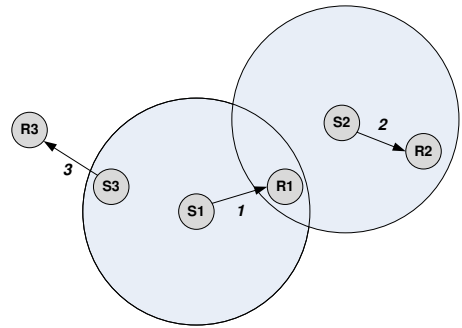}
\caption{Hidden and Exposed Terminal Problems} \label{fig: hidden_exposed}
\end{figure}

\subsection{Elimination of Hidden and Exposed Terminal Problems} \label{appen:hidden_exposed}

In IEEE 802.11 DCF, the RTS/CTS mechanism is used to reduce the \textbf{Hidden Terminal Problem}. However,
even if RTS/CTS is used, the hidden terminal problem can still occur, as illustrated in Fig.~\ref{fig:
hidden_exposed}, where we use $S_i$ and $R_i$ to denote the sender and receiver (in the sense of data
transmission) of link $i$. Under 802.11, it is possible that $S_2$ is sending an RTS to $R_2$ while $R_1$ is
returning a CTS to $S_1$ at the same time. $R_1$ cannot detect the RTS from $S_2$ since it is transmitting.
Likewise, $S_2$ cannot detect the CTS from $R_2$. Therefore, both links 1 and 2 could be scheduled, which
causes a collision at $R_1$.

Our Q-CSMA algorithm, however, can resolve the hidden terminal problem because the RTD/CTD messages are
exchanged in a synchronized manner. Suppose $T_i$ is the backoff expiration time of node $S_i$.
\begin{itemize}
\item $T_1 < T_2$: During the $(T_1 + 1)$-th mini-slot, $S_1$ sends an RTD in the first sub-mini-slot, then
in the second sub-mini-slot, $R_1$ returns a CTD, and link $1$ will be included in the decision schedule.
Since this CTD is not intended for $S_2$, $S_2$ disables its role as a sender of a link in the decision
schedule, thus link $2$ will not be included in the decision schedule.

\item $T_1 = T_2$: During the $(T_1 + 1)$-th mini-slot , both $S_1$ and $S_2$ send an RTD in the first
sub-mini-slot. In this case, $R_1$ senses a collision of RTD and will not return a CTD. Thus, link $1$ will
not be included in the decision schedule, while link $2$ will be included in the decision schedule.

\item $T_1 > T_2$: Similarly, in this case link $2$ but not link $1$ will be included in the decision
schedule.
\end{itemize}
Therefore, under synchronized RTD/CTD, the decision schedule is collision-free, which implies that the
transmission schedule is collision-free if we start with a collision-free transmission schedule (e.g., the
empty schedule), see Lemma~\ref{lemma:feasible}. Hence the hidden terminal problem is eliminated.

Another problem, known as the \textbf{Exposed Terminal Problem}, may also occurs under 802.11. In
Fig.~\ref{fig: hidden_exposed}, if $S_1$ sends an RTS to $R_1$, $S_3$ will receive this RTS and will be
silenced under 802.11, which is unnecessary because the potential transmission of link $3$ will not interfere
with link $1$. On the other hand, under Q-CSMA, if $S_1$ sends an RTD to $R_1$, $S_3$ will ignore this RTD
and can still send an RTD to $R_3$. Therefore, both links $1$ and $3$ can be included in the decision
schedule and in the transmission schedule, thus avoiding the exposed terminal problem.

Note that the presence of hidden and exposed terminals not only leads to loss of efficiency, but also poses
mathematical difficulties. For example, when there are hidden and exposed terminals in an 802.11-type
asynchronous RTS/CTS model, it is impossible to define a set of schedules that are consistent with both the
definition of a feasible schedule as defined in (\ref{eq:feasibleschedule}) and the capacity region as
defined in (\ref{eq:capacity}). Our RTD/CTD mechanism eliminates such problems.

\bibliographystyle{abbrv}
\bibliography{qcsma}

\begin{thebibliography}{10}

\bibitem{akyetal08}
U.~Akyol, M.~Andrews, P.~Gupta, J.~Hobby, I.~Saniee, and A.~Stolyar.
\newblock Joint scheduling and congestion control in mobile ad-hoc networks.
\newblock In {\em Proceedings of IEEE INFOCOM}, April 2008.

\bibitem{MACAW}
V.~Bharghavan, A.~Demers, S.~Shenker, and L.~Zhang.
\newblock {MACAW}: a media access protocol for wireless {LAN}'s.
\newblock In {\em Proceedings of ACM SIGCOMM}, 1994.

\bibitem{bia00}
G.~Bianchi.
\newblock Performance analysis of the {IEEE} 802.11 distributed coordination
  function.
\newblock {\em IEEE Journal on Selected Areas in Communications},
  18(3):535--547, 2000.

\bibitem{boo87}
R.~R. Boorstyn, A.~Kershenbaum, B.~Maglaris, and V.~Sahin.
\newblock Throughput analysis in multihop {CSMA} packet radio networks.
\newblock {\em IEEE Transactions on Communications}, 35(3):267--274, March
  1987.

\bibitem{bormcdpro08}
C.~Bordenave, D.~McDonald, and A.~Proutiere.
\newblock Performance of random multi-access algorithms, an asymptotic
  approach.
\newblock In {\em Proceedings of ACM Sigmetrics}, June 2008.

\bibitem{chakarsar05}
P.~Chaporkar, K.~Kar, and S.~Sarkar.
\newblock Throughput guarantees through maximal scheduling in wireless
  networks.
\newblock In {\em Proceedings of 43rd Annual Allerton Conference on
  Communication, Control and Computing}, 2005.

\bibitem{dimwal06}
A.~Dimakis and J.~Walrand.
\newblock Sufficient conditions for stability of longest-queue-first
  scheduling: Second-order properties using fluid limits.
\newblock {\em Advances in Applied Probabilities}, 38(2):505--521, 2006.

\bibitem{durthi06}
M.~Durvy and P.~Thiran.
\newblock Packing approach to compare slotted and non-slotted medium access
  control.
\newblock In {\em Proceedings of IEEE INFOCOM}, April 2006.

\bibitem{erysriper05}
A.~Eryilmaz, R.~Srikant, and J.~R. Perkins.
\newblock Stable scheduling policies for fading wireless channels.
\newblock {\em IEEE/ACM Transactions on Networking}, 13(2):411--424, April
  2005.

\bibitem{geoneetas06}
L.~Georgiadis, M.~Neely, and L.~Tassiulas.
\newblock Resource allocation and cross-layer control in wireless networks.
\newblock {\em Foundations and Trends in Networking}, 2006.

\bibitem{jiawal08b}
L.~Jiang and J.~Walrand.
\newblock Convergence analysis of a distributed {CSMA} algorithm for maximal
  throughput in a general class of networks.
\newblock {\em Technical Report, UC Berkeley}, December 2008.

\bibitem{jiawal08}
L.~Jiang and J.~Walrand.
\newblock A distributed {CSMA} algorithm for throughput and utility
  maximization in wireless networks.
\newblock In {\em Proceedings 46th Annual Allerton Conference on Communication,
  Control and Computing}, September 2008.

\bibitem{jiawal09}
L.~Jiang and J.~Walrand.
\newblock Approaching throughput-optimality in a distributed {CSMA} algorithm
  with contention resolution.
\newblock {\em Technical Report, UC Berkeley}, March 2009.

\bibitem{joolinshr08}
C.~Joo, X.~Lin, and N.~B. Shroff.
\newblock Understanding the capacity region of the greedy maximal scheduling
  algorithm in multi-hop wireless networks.
\newblock In {\em Proceedings of IEEE INFOCOM}, April 2008.

\bibitem{kel79}
F.~Kelly.
\newblock {\em Reversibility and Stochastic Networks}.
\newblock Wiley, Chichester, 1979.

\bibitem{lecnisri09}
M.~Leconte, J.~Ni, and R.~Srikant.
\newblock Improved bounds on the throughput efficiency of greedy maximal
  scheduling in wireless networks.
\newblock In {\em Proceedings of ACM MOBIHOC}, May 2009.

\bibitem{liekaileuwon07}
S.~C. Liew, C.~Kai, J.~Leung, and B.~Wong.
\newblock Back-of-the-envelope computation of throughput distributions in
  {CSMA} wireless networks.
\newblock Submitted for publication, http://arxiv.org//pdf/0712.1854.

\bibitem{linshrsri06}
X.~Lin, N.~B. Shroff, and R.~Srikant.
\newblock A tutorial on cross-layer optimization in wireless networks.
\newblock {\em IEEE Journal on Selected Areas in Communications}, 2006.

\bibitem{liuyipro08}
J.~Liu, Y.~Yi, A.~Proutiere, M.~Chiang, and H.~V. Poor.
\newblock Maximizing utility via random access without message passing.
\newblock {\em Microsoft Research Technical Report}, September 2008.

\bibitem{mareryozd07}
P.~Marbach, A.~Eryilmaz, and A.~Ozdaglar.
\newblock Achievable rate region of {CSMA} schedulers in wireless networks with
  primary interference constraints.
\newblock In {\em Proceedings of IEEE CDC}, December 2007.

\bibitem{mar99}
F.~Martinelli.
\newblock Lectures on {Glauber} dynamics for discrete spin models.
\newblock Lectures on probability theory and statistics (Saint-Flour, 1997),
  Lecture Notes in Math., 1717, Springer, Berlin, 1999.

\bibitem{proyichi08}
A.~Proutiere, Y.~Yi, and M.~Chiang.
\newblock Throughput of random access without message passing.
\newblock In {\em Proceedings of CISS}, March 2008.

\bibitem{rajsha08}
S.~Rajagopalan and D.~Shah.
\newblock Distributed algorithm and reversible network.
\newblock In {\em Proceedings of CISS}, March 2008.

\bibitem{rajshashi08}
S.~Rajagopalan, D.~Shah, and J.~Shin.
\newblock Aloha that works, November 2008.
\newblock Submitted.

\bibitem{shasri07}
S.~Shakkottai and R.~Srikant.
\newblock Network optimization and control.
\newblock {\em Foundations and Trends in Networking}, pages 271--379, 2007.

\bibitem{taseph92}
L.~Tassiulas and A.~Ephremides.
\newblock Stability properties of constrained queueing systems and scheduling
  policies for maximal throughput in multihop radio networks.
\newblock {\em IEEE Transactions on Automatic Control}, 37(12):1936--1948,
  December 1992.

\bibitem{wankar05}
X.~Wang and K.~Kar.
\newblock Throughput modelling and fairness issues in {CSMA/CA} based ad-hoc
  networks.
\newblock In {\em Proceedings of IEEE INFOCOM}, March 2005.

\bibitem{warjanrhe09}
A.~Warrier, S.~Janakiraman, and I.~Rhee.
\newblock {DiffQ}: Practical differential backlog congestion control for
  wireless networks.
\newblock In {\em Proceedings of IEEE INFOCOM}, 2009.

\bibitem{wusriper07}
X.~Wu, R.~Srikant, and J.~R. Perkins.
\newblock Queue-length stability of maximal greedy schedules in wireless
  networks.
\newblock {\em IEEE Transactions on Mobile Computing}, pages 595--605, June
  2007.

\bibitem{zusbrzmod08}
G.~Zussman, A.~Brzezinski, and E.~Modiano.
\newblock Multihop local pooling for distributed throughput maximization in
  wireless networks.
\newblock In {\em Proceedings of IEEE INFOCOM}, April 2008.

\end{thebibliography}

\end{document}